\documentclass[12pt]{article}
\usepackage[utf8]{inputenc}
\usepackage[T1]{fontenc}
\usepackage{mathptmx} 
\usepackage{amsmath,amssymb,amsfonts}
\usepackage{booktabs}
\usepackage{graphicx}
\usepackage{fancyhdr}
\usepackage{setspace}
\usepackage{geometry}
\usepackage{authblk}
\usepackage{url}
\usepackage{xcolor}
\usepackage{orcidlink}

\usepackage{authblk} 

\AtBeginDocument{%
  \renewcommand{\hbar}{\mathchar'26\mkern-9mu h}%
}

\hypersetup{pdfborder = {0 0 0}}

\title{\textbf{\Large{Reduced Gibbs free energy supply hinders brain information processing during mental fatigue}}}

\author[1,*]{\textbf{Danko D. Georgiev} \orcidlink{0000-0001-6846-1194}}

\author[1]{\textbf{Oskan B. Tasinov} \orcidlink{0000-0001-7268-3937}}

\author[1]{\textbf{Danail V. Pavlov} \orcidlink{0000-0001-7382-2054}}

\author[2]{\textbf{Deyan S. Hrusafov} \orcidlink{0009-0000-8167-6405}}

\affil[1]{Department of Biochemistry, Molecular Medicine and Nutrigenomics, Medical University of Varna, Varna, Bulgaria}

\affil[2]{Department of Psychiatry and Medical Psychology, Medical University of Varna, Varna, Bulgaria}

\affil[*]{Corresponding author: \href{mailto:danko.georgiev@mu-varna.bg}{danko.georgiev@mu-varna.bg}}

\date{}

\begin{document}

\maketitle
\thispagestyle{empty}

\vspace{-1.5cm}

\begin{center}
\setlength{\fboxsep}{12pt}
\setlength{\fboxrule}{0.6pt}
\fbox{%
\begin{minipage}{\dimexpr\textwidth-24pt\relax}
\small
\textbf{ABSTRACT}

\medskip
\noindent
\textbf{Background:} Brain information processing deteriorates as cortical neurons gradually transition from a rested state into fatigue. Subjectively, fatigue is experienced as a state of weariness, tiredness, or lack of energy that reduces the ability to work safely and effectively.

\smallskip
\noindent
\textbf{Objective:} In this theoretical paper, we pinpoint the physical origin of brain fatigue in the gradual deterioration of biochemical reaction quotients and transmembrane ion concentration gradients, which increase neuronal excitability and decrease the signal-to-noise ratio in the brain cortex.

\smallskip
\noindent
\textbf{Methods:} Brain performance in a rested state versus fatigue is examined by well-established, data-driven computer models for energy transport inside protein $\alpha$-helices in the presence of thermal noise for different ATP energy states or for pyramidal neuron firing of action potentials under electric stimulation in a rested membrane state versus fatigue.

\smallskip
\noindent
\textbf{Results:} We found that reduced Gibbs free energy supply from ATP hydrolysis impairs the cooperative effect between amide I excitons propagating inside protein $\alpha$-helices, with resulting decreased thermal stability of molecular solitons. Concurrent changes in Nernst reversal potentials for $\text{Na}^+$ or $\text{K}^+$ ions further led to neuronal hyperexcitability and a higher risk of neuronal depolarization block during mental fatigue.

\smallskip
\noindent
\textbf{Conclusions:} Detailed computational modeling showed that inefficient protein function due to diminished ATP energy status, alters the electrophysiological properties of individual neurons, thereby impairing their information processing capacity for the proper execution of cognitive tasks. Scheduling practices aimed at intermittent recovery of the rested brain state during intellectually challenging work could protect physical and mental wellbeing, prevent burnout, and enhance long-term productivity.


\medskip
\noindent
\textbf{Published in:} Fatigue: Biomedicine, Health \& Behavior 2026; 14 (3): 216--233.\\
DOI:~\href{http://doi.org/10.1080/21641846.2026.2636450}{10.1080/21641846.2026.2636450}

\end{minipage}%
}
\end{center}

\bigskip

\hypersetup{pdfborder = {1 1 1}}

\section{Introduction}

\begin{figure}[t!]
\begin{centering}
\includegraphics[width=\textwidth]{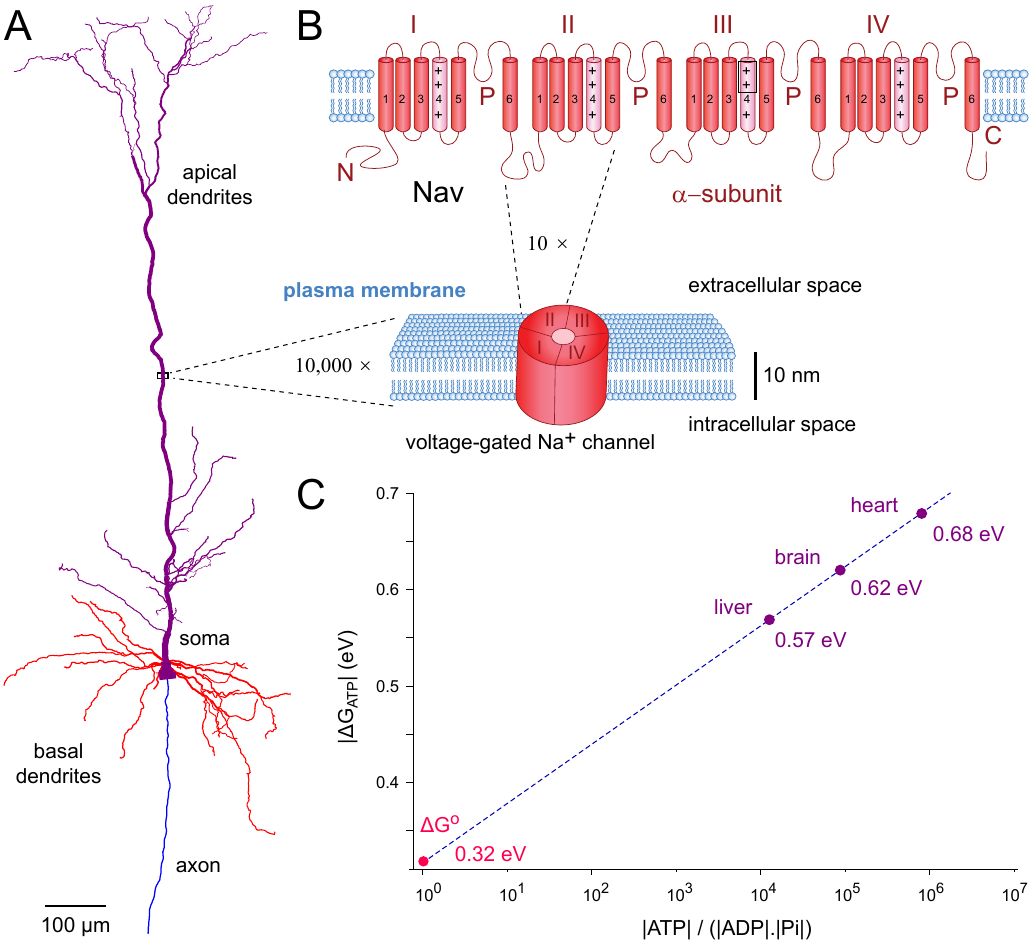}
\par\end{centering}
\caption{ATP utilization in the brain. (A) The morphology of a pyramidal neuron (NeuroMorpho~ID:~\href{https://neuromorpho.org/neuron_info.jsp?neuron_id=NMO_05726}{NMO\_05726}) from cortical layer 5 \cite{ref80}. (B) Voltage-gated $\text{Na}^+$ ion channels ($\text{Na}_{\text{v}}$) incorporated into the neuronal plasma membrane trigger the generation of action potentials upon electric stimulation. $\text{Na}_{\text{v}}$ $\alpha$-subunit forms the channel pore and consists of four repeat domains, labeled I through IV, each containing six protein $\alpha$-helices traversing the plasma membrane. The 4th protein $\alpha$-helix in each domain is positively charged and acts as a voltage sensor. ATP consumption by the neuronal $\text{Na}^+/\text{K}^+$ pump maintains the physiological $\text{Na}^+$ and $\text{K}^+$ ion concentration gradients, whose potential energy is used for electric signaling. (C) The amount of free energy $|\Delta G_{\text{ATP}}|$ released by ATP hydrolysis is higher in the brain and the heart, compared to the liver.}
\label{fig:1}
\end{figure}

The human brain is composed of 86 billion nerve cells called \emph{neurons} (Figure~\ref{fig:1}A), which input, process and output biologically relevant information in the form of electrical signals \cite{ref1,ref2}. In neuronal dendrites and soma, the electrical potentials are attenuated as they propagate in the form of either excitatory postsynaptic potentials (EPSPs) \cite{ref3} or inhibitory postsynaptic potentials (\mbox{IPSPs}) \cite{ref4}. The postsynaptic potentials are able to summate both spatially and temporally, thereby implementing a variety of computational logical gates \cite{ref5,ref6,ref7}. If the outcome of the implemented computational gate is above a certain threshold around $-55~\text{mV}$ of the membrane potential in the axonal hillock, the neuron fires an action potential \cite{ref8}. In axons, the action potentials are propagated without attenuation at significant distances where they trigger physiological responses in target cells, including neurons, muscle fibers or glands \cite{ref9,ref10}. At the molecular level, the electrical signals are generated by the opening or closing of different types of ion channels incorporated in the neuronal plasma membrane \cite{ref11,ref12,ref13,ref14}. Most of the ion channels are gated by changes in the transmembrane voltage and selectively transmit just a single ion type, such as $\text{Na}^+$, $\text{K}^+$ or $\text{Ca}^{2+}$ (Figure~\ref{fig:1}B) \cite{ref15,ref16}. The energy needed for sustaining the selective ion currents is provided by the corresponding ionic concentration gradients across the plasma membrane, and ultimately by hydrolysis of adenosine triphosphate (ATP) needed for the operation of the neuronal $\text{Na}^+/\text{K}^+$ pump \cite{ref17} or $\text{Ca}^{2+}$ pump \cite{ref18}.

The human brain has an average mass of $1508.91 \pm 299.14~\text{g}$ \cite{ref19} and utilizes glucose at a rate of $310 \pm 30~\text{nmol/g/min}$ \cite{ref20} to support its information processing activity. Aerobic catabolism of each glucose molecule delivers 38 ATP molecules \cite{ref21}. Thus, the brain consumes 25.6~mol ATP per day, being one of the most energetically expensive organs in the body. The other energy-demanding organ is the human heart with an average mass of 300~g and daily consumption of 9.86~mol ATP \cite{ref22}. Given that the ATP molar mass is $507.18~\text{g/mol}$, it can be estimated that the daily use of ATP by the heart is 5~kg, whereas by the brain it is 13~kg. The amount of ATP used, however, is a poor substitute for the actual free energy converted into useful work by biological systems. Even though, ATP is widely considered the `energy currency' of the cell, this metaphor should not be taken literally because the Gibbs free energy released by ATP hydrolysis $|\Delta G_{\text{ATP}}|$ is highly variable and strongly depends on the molar concentrations of ATP, adenosine diphosphate (ADP) and inorganic phosphate ($\text{P}_{\text{i}}$) inside the cells. This means that the same intracellular concentration of ATP can deliver different amounts of free energy $|\Delta G_{\text{ATP}}|$ leading to neuronal weakness and \emph{fatigue} in circumstances that prevent the cells from maintaining the appropriate reaction quotients far from thermodynamic equilibrium.

In this theoretical work, we study \emph{mental fatigue} as a reversible physiological state of submaximal cognitive performance by healthy subjects who do not have an accompanying neuropsychiatric disease. We utilize previously reported intracellular concentrations of different metabolites to compute the Gibbs free energy supplied by ATP hydrolysis in different organs, including the liver, brain and heart. Comparison of brain function in a rested state versus fatigue is then performed by inserting either the full amount of ATP energy or only a fraction of it into detailed, biologically accurate computer models for energy transport inside protein $\alpha$-helices in the presence of thermal noise \cite{ref23} or for pyramidal neuron firing action potentials under electric stimulation \cite{ref24}. We also quantify the trade-off between \emph{energy efficiency} and \emph{reliability of signal transmission} inside individual neuronal proteins and explain how degradation of the energy safety margin for proper protein function accumulates in the form of diminished ion concentration gradients, shifted Nernst reversal potentials and neuronal hyperexcitability. This provides a thermodynamic description of \emph{mental fatigue} as a reversible continuum of functional states that are bound between the two extremes of a fully rested neuronal state, which has maximal information processing capacity for cognitive tasks, and a fully exhausted neuronal state, which has no information processing capacity as manifested through cognitive paralysis due to neuronal depolarization block. The thermodynamic approach to mental fatigue exposes the existence of cognitive states of \emph{subtle incapacitation} during which the subject may lack any feeling of mental tiredness and is unaware of sporadic errors appearing in the execution of cognitive tasks.

\section{Methods}

Brain performance in a rested state versus fatigue is examined by physical thermodynamic equations and well-established, data-driven computer models for energy transport inside protein \mbox{$\alpha$-helices} or for pyramidal neuron firing of action potentials under electric stimulation.

\subsection{Gibbs free energy}

The Gibbs free energy equation relates the change in Gibbs free energy $\Delta G$ measured in electronvolts (eV) to the reaction quotient $Q$ of a chemical reaction:
\begin{equation}
\Delta G = \Delta G^{\circ} + \frac{k_{\text{B}} T}{q_{\text{e}}} \ln Q
\end{equation}
where $k_{\text{B}} = 1.380649 \times 10^{-23}~\text{J K}^{-1}$ is the Boltzmann constant, $q_{\text{e}} = 1.602176634 \times 10^{-19}~\text{C}$ is the elementary charge, $T$ is the absolute temperature in kelvin (K), $\Delta G^{\circ}$ is the standard change in Gibbs free energy when the concentrations of all aqueous solutions are 1 molar (M) at a temperature of $25^{\circ}~\text{C}$ ($T = 298.15~\text{K}$), and $Q$ is the reaction quotient computed from the thermodynamic activities of the substrates and products of the chemical reaction. $\Delta G$ is negative for spontaneous reactions, meaning that the conversion of substrates to products will proceed on its own under the given conditions.

For ATP hydrolysis, taking place inside neuronal cytosol,
\begin{equation}
\text{ATP} + \text{H}_2\text{O} \to \text{ADP} + \text{P}_{\text{i}}
\end{equation}
the reaction quotient is
\begin{equation}
Q = \frac{|\text{ADP}| \cdot |\text{P}_{\text{i}}|}{|\text{ATP}|}
\end{equation}
where the dimensionless thermodynamic activities of $|\text{ATP}|$, $|\text{ADP}|$ and $|\text{P}_{\text{i}}|$ are obtained by the corresponding intracellular molar concentrations $[\text{ATP}]$, $[\text{ADP}]$ and $[\text{P}_{\text{i}}]$ normalized by the standard concentration of $[1~\text{M}]$, namely,
\begin{equation}
|\text{ATP}| = \frac{[\text{ATP}]}{[1~\text{M}]}; \quad |\text{ADP}| = \frac{[\text{ADP}]}{[1~\text{M}]}; \quad |\text{P}_{\text{i}}| = \frac{[\text{P}_{\text{i}}]}{[1~\text{M}]}
\end{equation}
For the inverse reaction of ATP synthesis, the opposite change in Gibbs free energy is calculated from the inverse reaction quotient, $Q^{-1} = |\text{ATP}| / (|\text{ADP}| \cdot |\text{P}_{\text{i}}|)$, which determines the phosphorylation potential $\Delta G_{\text{P}} = -\Delta G_{\text{ATP}}$.

At thermodynamic equilibrium, there is no change in Gibbs free energy, $\Delta G = 0$, which implies that,
\begin{equation}
\Delta G^{\circ} = -\frac{k_{\text{B}} T}{q_{\text{e}}} \ln K_{\text{eq}}
\end{equation}
where $K_{\text{eq}}$ is the equilibrium constant of the chemical reaction. Since exergonic reactions release energy into the environment, $\Delta G^{\circ} < 0$ \cite{ref25}, they have equilibrium constants $K_{\text{eq}} > 1$. The cells utilize ATP as an energy source, because ATP hydrolysis has large negative $\Delta G^{\circ} = -0.32~\text{eV}$ compared to normoergic compounds.

Utilizing the mathematical properties of the logarithmic function to combine eqs. (1) and (5), the Gibbs free energy equation can be expressed in a compact form as,
\begin{equation}
\Delta G = \frac{k_{\text{B}} T}{q_{\text{e}}} \ln \frac{Q}{K_{\text{eq}}}
\end{equation}
When $Q < K_{\text{eq}}$, there are proportionally fewer products and more substrates present than at equilibrium, the reaction will tend to shift towards the right, favoring the forward reaction that consumes substrates in order to produce more products. The further from equilibrium are shifted the intracellular concentrations $[\text{ATP}]$, $[\text{ADP}]$ and $[\text{P}_{\text{i}}]$, the greater the amount of Gibbs free energy from ATP hydrolysis, $|\Delta G_{\text{ATP}}|$, which is made available for useful biological work.

The importance of different reaction quotients of ATP hydrolysis in different organs, is illustrated by the fact that metabolically more active organs sustain larger values of $|\Delta G_{\text{ATP}}|$, which is almost twice the absolute value of $\Delta G^{\circ}$. The energy released by ATP hydrolysis increases from liver, to brain, to heart (Figure~\ref{fig:1}C), predominantly due to maintenance of increasingly higher intracellular $[\text{ATP}]$ and lower intracellular $[\text{P}_{\text{i}}]$ concentrations (Table~\ref{tab:1}).

\begin{table}[t!]
\centering
\caption{Intracellular concentrations of ATP, ADP and $\text{P}_{\text{i}}$ in different organs.}
\label{tab:1}
\begin{tabular}{lccccc}
\toprule
\textbf{Organ} & \textbf{[ATP] (mM)} & \textbf{[ADP] (mM)} & \textbf{[$\text{P}_{\text{i}}$] (mM)} & \textbf{$|\Delta G_{\text{ATP}}|$ (eV)} & \textbf{Ref.} \\
\midrule
Liver & 2.72 & 0.061 & 3.45 & 0.57 & \cite{ref33} \\
Brain & 3.00 & 0.020 & 1.70 & 0.62 & \cite{ref34} \\
Heart & 9.66 & 0.042 & 0.29 & 0.68 & \cite{ref35} \\
\bottomrule
\end{tabular}
\end{table}

\subsection{Davydov's model of molecular solitons}

The quantum equations of motion that govern the propagation of amide I excitations in Davydov's model of protein $\alpha$-helix have been previously derived by a number of research teams \cite{ref26,ref27}. In the presence of a thermal bath with temperature $T = 310~\text{K}$, the system of stochastic Langevin equations of motion for the full three-spine model of a protein $\alpha$-helix is,
\begin{equation}
i ~\hbar \frac{d}{dt} a_{n,\alpha} = -J_1(a_{n+1,\alpha} + a_{n-1,\alpha}) + J_2(a_{n,\alpha+1} + a_{n,\alpha-1}) + \chi(b_{n+1,\alpha} - b_{n-1,\alpha})a_{n,\alpha}
\end{equation}
\begin{align}
M \frac{d^2}{dt^2} b_{n,\alpha} = & w_1(b_{n-1,\alpha} - 2b_{n,\alpha} + b_{n+1,\alpha}) + w_2(b_{n,\alpha-1} - 2b_{n,\alpha} + b_{n,\alpha+1}) \nonumber \\
& - \Lambda \chi (|a_{n-1,\alpha}|^2 - |a_{n+1,\alpha}|^2) - M \Gamma \frac{d}{dt} b_{n,\alpha} + \sqrt{2M\Gamma k_{\text{B}} T} \frac{d}{dt} W_{n,\alpha}(t)
\end{align}
where the index $n = \{1, 2, \dots, 40\}$ counts the peptide groups along each protein $\alpha$-helix spine, the modular index $\alpha = \{0, 1, 2\}$ counts the three $\alpha$-helix spines modulo 3, $a_{n,\alpha}$ are the complex-valued amide I quantum probability amplitudes at the peptide group $n, \alpha$; $\Lambda$ is the number of amide I excitons, $b_{n,\alpha}$ is the expectation value of the displacement operator from the equilibrium position of the peptide group $n, \alpha$; $J_1 = 967.4~\mu\text{eV}$ is the nearest neighbor dipole-dipole coupling energy along the spine, $J_2 = 1535.4~\mu\text{eV}$ is the nearest lateral neighbor dipole-dipole coupling energy between different spines, $w_1 = 13~\text{N/m}$ is the spring constant of the longitudinal hydrogen bonds in the lattice of peptide groups, $w_2 = 30.5~\text{N/m}$ is the spring constant of the lateral coupling of the lattice of peptide groups due to covalent bonds in the protein backbone, $\chi = 35~\text{pN}$ is a nonlinear coupling parameter that parameterizes the interaction strength between the amide I excitons and the peptide group lattice displacements, $M = 1.9 \times 10^{-25}~\text{kg}$ is the average mass of an amino acid, $\Gamma = 0.005 \sqrt{w_1/M}$~Hz is the lowest (non-zero) frequency of the protein lattice, and $W_{n,\alpha}(t)$ are independent real-valued continuous-time stochastic Wiener processes with zero drift and unit volatility \cite{ref23}. The inclusion of the damping term $- M \Gamma \frac{d}{dt} b_{n,\alpha}$ and a thermal noise term $\sqrt{2M\Gamma k_{\text{B}} T} \frac{d}{dt} W_{n,\alpha}(t)$, which obey the fluctuation-dissipation theorem \cite{ref28}, ensures that on picosecond timescale the average temperature of the protein $\alpha$-helix is maintained at $T = 310~\text{K}$.

The initial Gaussian distribution of amide I quantum probability amplitudes was non-zero only over the first five peptide groups inside a 18-nm long protein $\alpha$-helix as described previously \cite{ref29}:
\begin{align*}
a_{1,\alpha} & = \sqrt{0.099} \exp\left( -i\frac{\pi}{2} + i\frac{\pi}{3}\alpha \right), \nonumber \\
a_{2,\alpha} & = \sqrt{0.24} \exp\left( -i\frac{\pi}{4} + i\frac{\pi}{3}\alpha \right), \nonumber \\
a_{3,\alpha} & = \sqrt{0.322} \exp\left( i\frac{\pi}{3}\alpha \right), \nonumber \\
a_{4,\alpha} & = \sqrt{0.24} \exp\left( i\frac{\pi}{4} + i\frac{\pi}{3}\alpha \right), \nonumber \\
a_{5,\alpha} & = \sqrt{0.099} \exp\left( i\frac{\pi}{2} + i\frac{\pi}{3}\alpha \right)
\end{align*}
For rigorous comparison of quantum dynamics with different number of amide I quanta $\Lambda = 3$ vs $\Lambda = 2$, the same set of 120 randomly generated Wiener processes $W_{n,\alpha}(t)$ was injected as thermal noise in both simulations.

\subsection{Electrophysiological model of a CA1 pyramidal neuron}

A morphologically complete model of CA1 pyramidal neuron was simulated with NEURON~8.2.7 (\url{https://neuron.yale.edu/neuron/}) \cite{ref30}. The reconstructed 3D morphology, including soma, apical dendrites, basal dendrites and axon, was based on NeuroMorpho ID:~\href{https://neuromorpho.org/neuron_info.jsp?neuron_id=NMO_00123}{NMO\_00123} \cite{ref31}. The model contained two types of voltage-gated $\text{Na}^+$ channels generating transient $\text{Na}^+$ current ($I_{\text{NaT}}$) and persistent $\text{Na}^+$ current ($I_{\text{NaP}}$); three types of voltage-gated $\text{K}^+$ channels generating delayed rectifier $\text{K}^+$ current ($I_{\text{KDR}}$), A-type $\text{K}^+$ current ($I_{\text{KA}}$) and M-type $\text{K}^+$ current ($I_{\text{KM}}$); three types of voltage-gated $\text{Ca}^{2+}$ channels generating T-type $\text{Ca}^{2+}$ current ($I_{\text{CaT}}$), R-type $\text{Ca}^{2+}$ current ($I_{\text{CaR}}$) and L-type $\text{Ca}^{2+}$ current ($I_{\text{CaL}}$); two types of $\text{Ca}^{2+}$-dependent $\text{K}^+$ channels generating slow afterhyperpolarization $\text{Ca}^{2+}$-dependent $\text{K}^+$ current ($I_{\text{sAHP}}$) and medium afterhyperpolarization $\text{Ca}^{2+}$-dependent $\text{K}^+$ current ($I_{\text{mAHP}}$); and mixed conductance hyperpolarization activated h-current ($I_{\text{h}}$) \cite{ref24}. Physiological ion channel densities in different neuronal compartments followed experimentally established distributions \cite{ref7,ref24} and are freely available online from ModelDB database (ModelDB ID:~\href{https://modeldb.science/143719}{143719}).

The driving force for the individual ionic currents $I_j$ through voltage-gated ion channels is determined by the specific conductance $g_j$ of the channel, and the difference between the actual voltage of the plasma membrane $V_{\text{m}}$ and the Nernst reversal potential $E_j$ for the particular type of ion $j$ \cite{ref32}:
\begin{equation}
I_j = g_j(V - E_j)
\end{equation}

The Nernst reversal potential depends on the valence $z$ of the ionic species $j$, as well as the ratio of the extracellular and intracellular ionic concentrations:
\begin{equation}
E_j = \frac{k_{\text{B}} T}{z q_{\text{e}}} \ln \frac{[j]_{\text{out}}}{[j]_{\text{in}}}
\end{equation}

In a rested neuronal state, $\text{Na}^+$ ions have a large positive reversal potential $E_{\text{Na}} = 71~\text{mV}$, due to their predominant extracellular localization, $[\text{Na}^+]_{\text{out}} = 145~\text{mM}$, $[\text{Na}^+]_{\text{in}} = 10~\text{mM}$, whereas $\text{K}^+$ ions have a large negative reversal potential, $E_{\text{K}} = -89~\text{mV}$, due to their predominant intracellular localization, $[\text{K}^+]_{\text{out}} = 5~\text{mM}$, $[\text{K}^+]_{\text{in}} = 140~\text{mM}$. To simulate physical conditions of \emph{mental fatigue}, we have investigated neuronal firing with degraded Nernst reversal potentials, $E_{\text{Na}} = 60~\text{mV}$ and $E_{\text{K}} = -70~\text{mV}$, and compared the performance with the one achieved in rested state, $E_{\text{Na}} = 71~\text{mV}$ and $E_{\text{K}} = -89~\text{mV}$.

\section{Results}

\subsection{Importance of ATP energy for protein function}

We found that in the rested state, the thermodynamically allowed Gibbs free energy released by a single ATP molecule is $0.57~\text{eV}$ in the liver \cite{ref33}, $0.62~\text{eV}$ in the brain \cite{ref34}, and $0.68~\text{eV}$ in the heart \cite{ref35}. The organs that have high energy demands achieve this large $|\Delta G_{\text{ATP}}|$ by shifting the reaction quotient for ATP hydrolysis far away from thermodynamic equilibrium. Different amounts of free ATP energy are able to excite different number of amide I exciton quanta inside protein $\alpha$-helices, where the cooperative effect between the excitons can significantly affect the protein functionality in the presence of thermal noise \cite{ref23}.

\begin{figure}[t!]
\begin{centering}
\includegraphics[width=\textwidth]{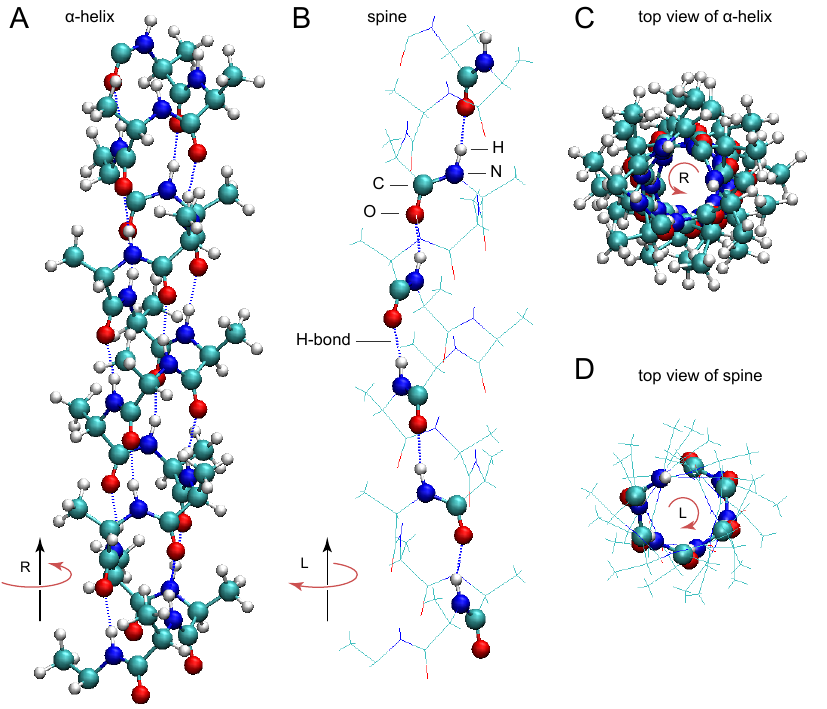}
\par\end{centering}
\caption{Atomic structure of poly-alanine protein $\alpha$-helix. (A) Side view of right-handed protein $\alpha$-helix with 3.6 amino acid residues per turn and translation of 1.5~\AA\ per amino acid along the helical axis. (B) The $\alpha$-helical structure is supported by 3 spines comprised of hydrogen bonds between the N-H group of one amino acid and the C=O group of the amino acid four residues earlier. The peptide groups (--CO--NH--) forming one of the spines are shown with full atomic radii. (C) Top view of the protein $\alpha$-helix. (D) Top view of the $\alpha$-helix spine.}
\label{fig:2}
\end{figure}

Efficient energy transport and utilization of exciton energy inside proteins is indispensable for their catalytic functions \cite{ref36}. The site of ATP hydrolysis could be located at a distance from the protein active site, which necessitates a physical mechanism for the transfer of the free energy to the active site where it is utilized to do useful work. If $0.62~\text{eV}$ of free energy were released from ATP in the form of an infrared photon, it would have had a wavelength $\lambda = 2~\mu\text{m}$, which is 400 times wider than the average protein diameter of $5~\text{nm}$ \cite{ref37}. The emission of such an infrared photon would have had a very low probability of being absorbed at the protein active site and could not have explained the high efficiency of protein function. The chemical composition of the polypeptide chains of a sequence of peptide bonds (--CO--NH--), however, allows for energy transport in the form of amide I (--CO--) and amide II band (--NH--) vibrations \cite{ref38}. It is noteworthy that individual amide I excitons have an energy of $0.2~\text{eV}$ and can propagate along the hydrogen-bonded peptide groups inside protein $\alpha$-helices (Figure~\ref{fig:2}) \cite{ref39,ref40}. Depending on the available amount of free ATP energy, two or three amide I excitons can be generated in the cells of different organs. Due to the existing cooperative effect among the amide I excitons, the greater number of exciton quanta stabilizes the traveling solitary wave (soliton), increases its lifetime and prevents its dispersal in the presence of thermal noise.

\begin{figure}[t!]
\begin{centering}
\includegraphics[width=\textwidth]{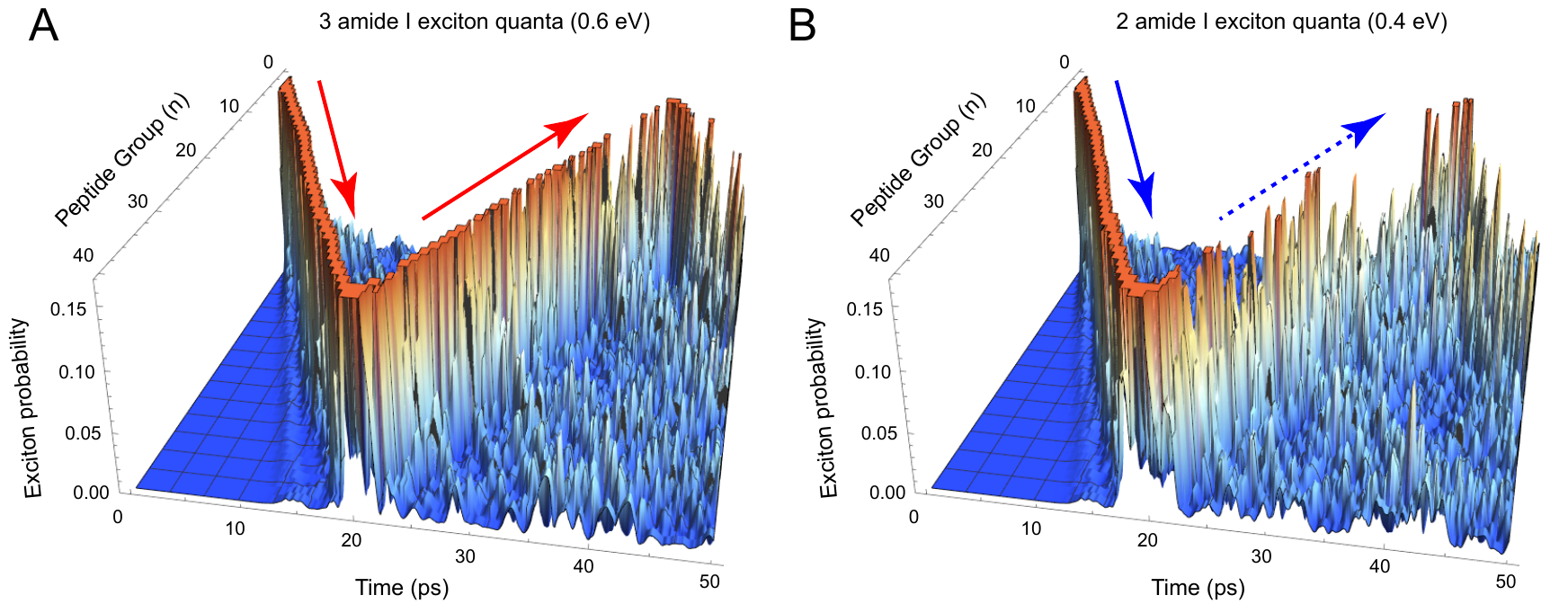}
\par\end{centering}
\caption{Thermal stability of protein solitons with different numbers of amide I exciton quanta at physiological temperature of $T = 310~\text{K}$. (A) Protein soliton with 3 amide I quanta ($0.6~\text{eV}$) moves at speed of 900~m/s and has a lifetime of at least 50~ps. Such soliton can deliver its energy for utilization at a distance of 45~nm away from the origin. (B) Protein soliton with 2 amide I quanta ($0.4~\text{eV}$) moves at speed of 938~m/s but it is dispersed by the thermal noise after $\approx 20$~ps. Such soliton can deliver its energy for utilization at a shorter distance less than 19~nm away from the origin. Solid arrows indicate the direction of soliton propagation. Dashed arrows indicate the effect of thermal dispersion.}
\label{fig:3}
\end{figure}

We performed detailed computer simulations based on Davydov's model of energy transport inside a protein $\alpha$-helix immersed in a thermal bath with physiological temperature $T = 310~\text{K}$ \cite{ref23}, which showed that a protein soliton comprised of $\Lambda = 3$ amide I exciton quanta has a lifetime of at least 50~ps (Figure~\ref{fig:3}A), whereas a soliton comprised of only $\Lambda = 2$ amide I exciton quanta persists for approximately 20~ps (Figure~\ref{fig:3}B). The significance of longer lifetime is that the soliton can deliver its energy for utilization in biological work at a longer distance from the ATP hydrolytic site. For $\Lambda = 3$, the maximal distance is up to 45~nm away, whereas for $\Lambda = 2$ it is less than 19~nm. Energy-demanding organs, such as the brain and the heart, are able to minimize thermal waste of ATP energy by tightly regulating the intracellular concentrations of $[\text{ATP}]$, $[\text{ADP}]$ and $[\text{P}_{\text{i}}]$. If the resulting reaction quotient given by eq. (3) supports at least $0.6~\text{eV}$ of free energy released per ATP molecule, the cellular proteins could utilize the stabilizing cooperative effect of 3 amide I exciton quanta in order to perform their mechanical or catalytic function.

The quantization of energy in quantum physics implies that the excitation of a single quantum requires a specific, minimum amount of energy $E$ dictated by the Planck--Einstein relation, $E = h \nu$, where $h = 6.62607015 \times 10^{-34}~\text{J/Hz}$ is the Planck constant and $\nu$ is the frequency of oscillation. If the supplied energy is less than this minimum, the quantum cannot be excited. In the rested state, $0.62~\text{eV}$ of ATP energy is sufficient to excite 3 amide I exciton quanta, however if $|\Delta G_{\text{ATP}}| < 0.6$ it will be able to excite at most 2 amide I exciton quanta. The discrete threshold for generation of amide I quanta allows for rigorous quantitative evaluation of the trade-off between \emph{energy efficiency} and \emph{reliability of signal transmission} inside individual neuronal proteins. In rested state, the energy efficiency of brain proteins is extremely high, $> 96.7\%$, with only $0.02~\text{eV}$ of ATP energy not converted into useful molecular soliton. This, however, implies that the safety margin for reliable transmission of the molecular soliton is very narrow, $< 3.3\%$. If the free energy from ATP hydrolysis drops below $0.6~\text{eV}$ as neuronal fatigue progresses, the maximal distance for reliable transmission of molecular solitons becomes restricted to $< 42.3\%$ of the rested range.

It is worth emphasizing that the spatial extent and complexity of neuronal arborizations (dendrites and axons) is at least 6 orders of magnitude larger than the biomolecular nanoscale \cite{ref1}. This means that the concentrations of ATP, ADP and $\text{P}_{\text{i}}$ are compartmentalized at the micrometer level \cite{ref41,ref42}, vary significantly between individual compartments such as individual dendritic spines \cite{ref43}, and the resulting $|\Delta G_{\text{ATP}}|$ yields should be extracted from the semilogarithmic plot shown in Figure~\ref{fig:1}C. Synaptic activity induces the recruitment and anchoring of mitochondria under active spines to locally supply the high levels of ATP required for synaptic plasticity \cite{ref44}. Local depletion of ATP can be temporarily counteracted by brain creatine kinase, which consumes compartmentalized depots of creatine phosphate \cite{ref45}. Repetitive neuronal stimulation from tens of seconds to minutes, significantly depletes ATP reserves and is manifested in failure of ionic pumps to maintain the resting transmembrane ionic gradients \cite{ref46,ref47}. This leads to progression of mental fatigue from the microscopic subcellular compartments to the cellular level.

\subsection{Conversion of ATP energy into ion concentration gradients}

Electrically active cells such as neurons and muscle cells, rely on the concerted opening and closure of a variety of selective ion channels \cite{ref13}. The passage of electric currents through the ion channels is powered by the ion concentration gradients existing between the extracellular matrix and the cytosol \cite{ref9}. The individual ionic Nernst reversal potentials are dynamically regulated by ionic pumps, which set the intracellular and extracellular ionic concentrations at the expense of ATP energy \cite{ref48}. Under normal physiological conditions, over $53\%$ of the neuronal ATP budget is consumed by the $\text{Na}^+/\text{K}^+$ pump for maintaining proper neuronal excitability \cite{ref1}. During intense electric activity of brain neurons, the ionic concentrations gradients diminish due to accumulation of $\text{Na}^+$ ions inside the neurons and leakage of $\text{K}^+$ ions to the extracellular space. This leads to lower absolute values of the ionic reversal potentials, lower driving force for the ionic currents and may eventually lead to neuronal depolarization block.

\begin{figure}[t!]
\begin{centering}
\includegraphics[width=\textwidth]{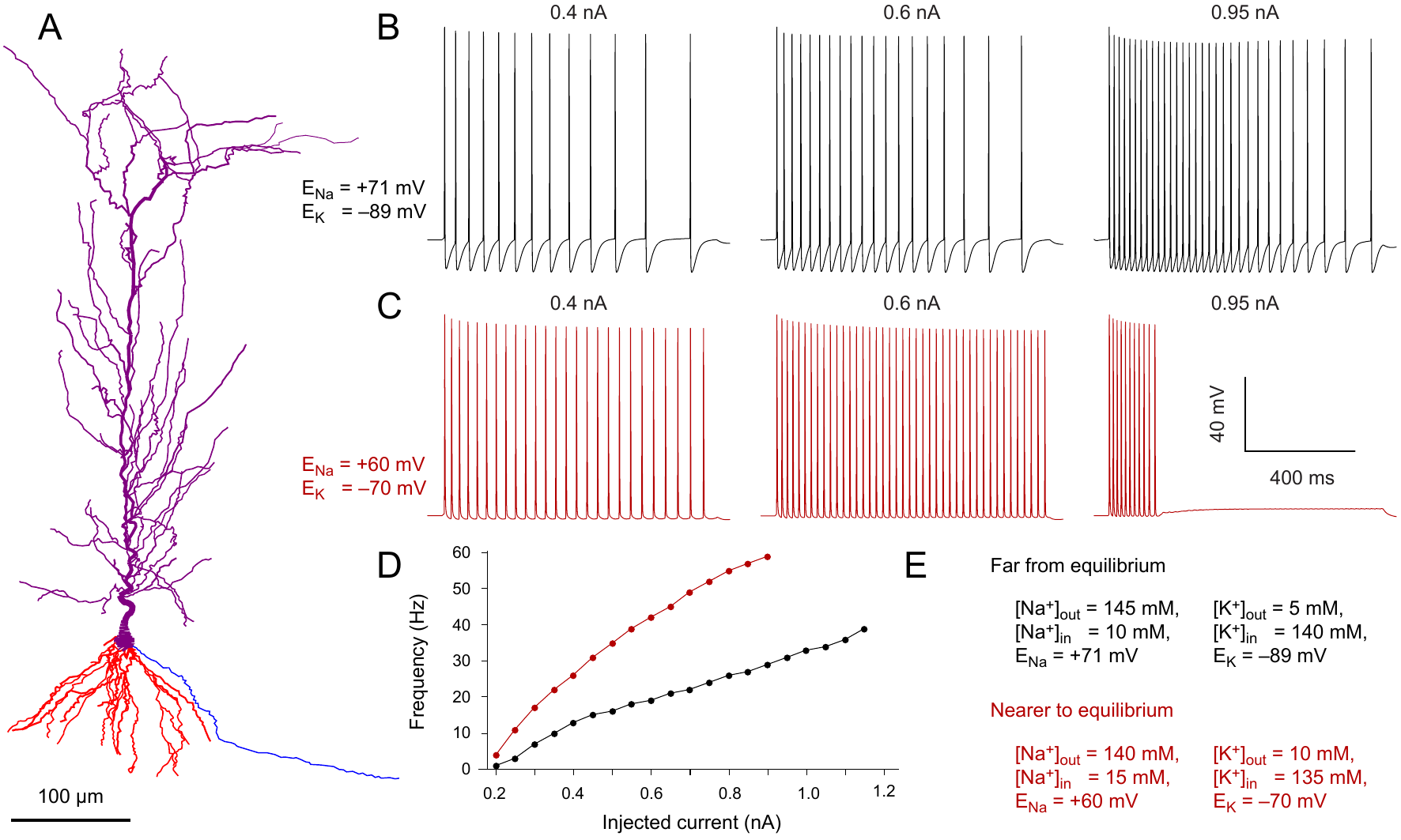}
\par\end{centering}
\caption{CA1 pyramidal neuron (NeuroMorpho ID:~\href{https://neuromorpho.org/neuron_info.jsp?neuron_id=NMO_00123}{NMO\_00123}) \cite{ref31} firing under different electrolyte conditions. (A) Computer simulation (ModelDB ID:~\href{https://modeldb.science/143719}{143719}) \cite{ref24} of electric activity for injected somatic currents of different magnitude (nA) under rested condition far from equilibrium, $E_{\text{Na}} = 71~\text{mV}$ and $E_{\text{K}} = -89~\text{mV}$. (C) Electric activity for injected somatic currents of different magnitude (nA) under fatigue condition nearer to equilibrium, $E_{\text{Na}} = 60~\text{mV}$ and $E_{\text{K}} = -70~\text{mV}$. (D) Comparison of f-I curves exhibiting the relationship between injected current (I) and firing frequency (f) under rested condition far from equilibrium (black) and fatigue condition nearer to equilibrium (red). (E) Intracellular and extracellular concentrations for $\text{Na}^+$ and $\text{K}^+$ ions, together with the Nernst reversal potentials $E_{\text{Na}}$ and $E_{\text{K}}$ for the two conditions under comparison.}
\label{fig:4}
\end{figure}

We performed detailed computer simulations of a morphologically complete CA1 pyramidal neuron (ModelDB ID:~\href{https://modeldb.science/143719}{143719}) \cite{ref24}, which revealed a large functional effect of Nernst reversal potentials upon neuronal electric firing in the brain (Figure~\ref{fig:4}). Under rested physiological conditions far from equilibrium, $E_{\text{Na}} = 71~\text{mV}$ and $E_{\text{K}} = -89~\text{mV}$, the CA1 pyramidal neuron exhibits adaptation manifested by widening interspike time intervals and decreasing firing frequency during prolonged excitation \cite{ref24} (Figure~\ref{fig:4}B). After prolonged mental work, fractions of the ion concentration gradients for $\text{Na}^+$ and $\text{K}^+$ ions are dissipated and the corresponding Nernst reversal potentials are shifted nearer to equilibrium. Electrophysiological recordings from rat hippocampus have shown that during repetitive electric stimulation with duration of 10~s, the extracellular $\text{K}^+$ concentration can easily surpass $[\text{K}^+]_{\text{out}} = 10~\text{mM}$ \cite{ref49}. During normal mental work, repetitive neuronal stimulation of the same brain region could be sustained for minutes. Conservatively, if only 5~mM of intracellular $\text{K}^+$ ions are exchanged for $\text{Na}^+$ ions, the reversal potentials shift to $E_{\text{Na}} = 60~\text{mV}$ and $E_{\text{K}} = -70~\text{mV}$, which makes the pyramidal neuron more excitable and prone to entering into a depolarization block (Figure~\ref{fig:4}C). Comparison of the f-I curves far from equilibrium or nearer to equilibrium (Figure~\ref{fig:4}D) shows that the pyramidal neuron can tolerate much higher levels of external excitation (quantified by the 1.15~nA injected current in the soma) maintaining physiological frequency of firing below 40~Hz as long as the neuronal $\text{Na}^+/\text{K}^+$ pump can restore the $\text{Na}^+$ and $\text{K}^+$ ion concentrations to their resting values far from equilibrium (Figure~\ref{fig:4}E). Besides the possibility of a depolarization block, the enhanced excitability of pyramidal neurons is detrimental to the brain processing of information due to decreased signal to noise ratio, with spontaneous firing or erratic responses to background noise as irrelevant fluctuations become more prominent relative to the desired sensory signal. When neural excitability is high, it can lead to a state of \emph{hyperarousal} where the brain is easily overwhelmed by sensory input, making it difficult to filter out distractions and maintain focus on the execution of a specific task \cite{ref50,ref51}. In fact, prevention of age-associated neuronal hyperexcitability improves attention, memory performance and learning \cite{ref52}.

Alterations in excitability of individual neurons accumulate in time and can be detected at the macroscopic level using electroencephalography (EEG) \cite{ref53}. EEG has low spatial resolution of 22--37~$\text{cm}^3$ for standard 19-electrode EEG systems and 6--8~$\text{cm}^3$ for high-density 129-electrode EEG systems \cite{ref54}. Because the human cortex has on average 20 million neurons per $\text{cm}^3$ \cite{ref19}, EEG can register mental fatigue when a macroscopic number of $10^7$--$10^8$ cortical neurons are electrically altered. This occurs relatively late in the progression of mental fatigue and usually requires mental workload that lasts 2--3 hours \cite{ref53,ref55}. Our simulation of an individual pyramidal neuron highlights that the thermodynamic origin of mental fatigue through degradation of Nernst ionic reversal potentials could affect cognitive processing of information even before there are macroscopic changes detectable through EEG. Alternatively, human functional magnetic resonance imaging (fMRI) may improve spatial resolution to voxels with $3 \times 3 \times 3~\text{mm}^3$ \cite{ref56}, however, the blood oxygenation level-dependent (BOLD) signal measures the brain activity indirectly by detecting changes in blood oxygen levels, relying on the fact that active neurons consume oxygen and trigger a localized increase in oxygen-rich blood flow, which alters the magnetic properties of hemoglobin, creating a detectable MRI signal change \cite{ref57}. Despite that the BOLD signal originates from erythrocytes in the blood vessels, instead from neurons, it can detect certain types of abnormal brain activity manifested through desynchronization or miscommunication between distinct brain networks \cite{ref58,ref59,ref60}. Interpretation of macroscopic brain activity in terms of underlying neuronal processing of information, however, heavily relies on data obtained from detailed computational models of individual neurons, such as the one reported in Figure~\ref{fig:4}, which pinpoint the precise physical origin of mental fatigue and allow for development of informed measures for prophylaxis.

\section{Discussion}

We found that reduced Gibbs free energy supply from ATP hydrolysis impairs significantly the information processing capacity of the brain through deterioration of protein molecular functions at the nanolevel and alteration of neuronal membrane excitability at the microscopic cellular level. The overarching physical thermodynamic mechanism underlying mental fatigue can be summarized in 4 steps: (1) accumulation of reaction products, (2) increased reaction quotient, (3) insufficient release of Gibbs free energy, and (4) impaired brain information processing performance with lower signal-to-noise ratio.

\subsection{Fatigue as an inefficient molecular function due to insufficient energy supply}

Fatigue is a symptom, subjectively described as an overall feeling of tiredness, weariness, lack of energy, or exhaustion \cite{ref61}. Slow response times and decline in accuracy during cognitive performance tasks can be an objective sign of mental fatigue \cite{ref62}. The subjective characterization of fatigue as a `lack of energy' is indeed conceptually very close to the thermodynamic inability of molecular systems to perform useful work. Psychological studies by O'Connor and Boolani \cite{ref63,ref64,ref65} suggest that the \emph{subjective feeling of fatigue} and \emph{metabolic energy deficit} are two separate constructs and that metabolic energy deficit may be associated with predictors of cognitive performance and not the feeling of mental fatigue. Our thermodynamic description of mental fatigue as a reversible continuum of functional states, which are bound between the two extremes of a fully rested neuronal state and a fully exhausted neuronal state, indeed supports such psychological distinction and provides a physical explanation for the existence of cognitive states of subtle incapacitation during which the subject may lack any feeling of mental tiredness and is unaware of sporadic errors appearing in the execution of cognitive tasks. In the particular case of the brain, decreased supply with ATP energy is expressed as an impairment of neuronal information processing capacity \cite{ref66}. The input, processing and transmission of biologically relevant information is achieved by neurons by keeping the intracellular concentrations of metabolites, ATP and electrolytes far from thermodynamic equilibrium \cite{ref34}. Physical states far from equilibrium release larger amounts of Gibbs free energy in shorter amount of time leading to cooperative effects and highly efficient utilization for biological work with minimized direct waste in the form of heat \cite{ref23}. In contrast, operating nearer to thermodynamic equilibrium is associated with lower amounts of Gibbs free energy released from ATP hydrolysis and only partial utilization of that energy (Figure~\ref{fig:3}) to support protein functions.

Mental fatigue is not a discrete on/off phenomenon, but a gradual physiological process that develops throughout the working day. The initial stages of mental fatigue do not need be associated with subjective feeling of tiredness and the cognitive performance could be diminished without awareness of the ongoing \emph{subtle incapacitation} \cite{ref67}. The feeling of being mentally tired could be a relatively late result from the conscious recognition of slow cognitive performance, registration of sporadic cognitive errors or the steady accumulation of extracellular glutamate in the prefrontal cortex \cite{ref68,ref69}. Glutamate is the major excitatory neurotransmitter in the neocortex, with $\approx 80\%$ of the cortical neurons being glutamatergic \cite{ref70}. Whereas basal glutamatergic neurotransmission triggers pro-survival biochemical pathways and release of brain-derived neurotrophic factor to support neurite growth and extension \cite{ref71}, prolonged glutamate release can lead to neuronal hyperexcitation and excitotoxicity \cite{ref72}. Cortical astroglial cells, typically take up the excess glutamate from the neuronal synapses through $\text{Na}^+/\text{glutamate}$ transporters, convert glutamate into glutamine and transport the glutamine back to neurons, where it acts as a main source of newly synthesized glutamate \cite{ref73,ref74}. Under rested conditions, this astroglial glutamate--glutamine shuttle metabolizes $\approx 80\%$ of glutamate released during synaptic transmission \cite{ref75}. Under prolonged brain activity, however, the accumulation of glutamine inside astrocytes decreases the efficiency of glutamate uptake and the synaptic glutamate is able to diffuse into the brain interstitial space \cite{ref76}, where its buildup causes subjectively reportable symptoms of mental fatigue, including brain fog, irritability, poor concentration, and a lack of motivation.

Macroscopic changes of brain metabolites, such as glutamate, glutamine or creatine phosphate could be measured through magnetic resonance spectroscopic imaging, however, they are mainly useful for registration of pronounced fatigue after daylong cognitive work when decision making is already compromised \cite{ref68}. For prevention of life-threatening errors due to subtle incapacitation, for example, in airplane pilots or air traffic controllers \cite{ref77}, it is necessary to act at early stages of mental fatigue, when the subject is still unaware of compromised cognitive decision making abilities.

\subsection{Prophylaxis of mental fatigue}

Brain transition from a rested state into a state of mental fatigue is a gradual process that shifts the brain molecular constituents nearer to thermodynamic equilibrium. The resulting changes in the reaction quotient of ATP hydrolysis, ionic concentration gradients and electromotive forces for different ion types, increase the excitability of brain neurons and impair their ability to differentiate between sensory signals and irrelevant noise. Because even a small drop in the supply of Gibbs free energy from ATP hydrolysis $|\Delta G_{\text{ATP}}|$ could drastically reduce the number of excited amide I quanta from 3 to 2 inside neuronal proteins, it may be justified to prefer shorter in duration but more frequent attempts at restoration of the initial rested brain state, as opposed to longer in duration but less frequent pauses for rest. Therefore, intermittent recovery of the rested brain state through scheduling practices for rest during mentally challenging work can protect physical and mental wellbeing, prevent burnout, and enhance long-term performance and productivity. Some Asian countries, such as China and Japan, encourage napping during work hours, because such practice has been found to promote alertness on the job \cite{ref78}.

Having elaborated on the concrete molecular and thermodynamic mechanisms, which substantiate the description of mental fatigue as a `lack of energy', it is important to emphasize the necessity of prophylaxis with proactive management and prevention of mental fatigue through focus on regular breaks, sufficient sleep, stress management techniques, and maintenance of a healthy lifestyle \cite{ref79}. As long as mental fatigue is reversible, it could be viewed as a part of the daily physiological process of brain energy depletion and replenishment. In this theoretical work, we have not included the effects of irreversible neuronal changes that occur in neuropsychiatric disease in order to better differentiate between reversible `fatigue' and irreversible neuron `injury'. Within nosology, mental fatigue is a core symptom of depression and anxiety disorders, and it is also prevalent in neurological illnesses such as multiple sclerosis and Parkinson's disease. In such cases, the clinician has to consider both the effects of neuronal fatigue and injury for making therapeutic decisions.

\section{Conclusions}

In this theoretical paper, we have investigated the thermodynamic origin of mental fatigue due to changes in intracellular metabolite concentrations and transmembrane ion concentration gradients, with resulting impaired supply of Gibbs free energy that directly compromises the capacity of the brain to do useful work. At the nanolevel, we have presented computational evidence of impaired transport and utilization of energy in proteins during fatigue due to decreased thermal stability of molecular solitons. At the cellular level, insufficient energy supply to the neuronal ion pumps shifts the Nernst reversal potentials for $\text{Na}^+$ and $\text{K}^+$ ions, increases neuronal excitability and leads to progressive accumulation of extracellular glutamate in the brain cortex as the fatigue increases. The impaired information processing by the brain cortex is unfolded gradually as the metabolic and ionic concentrations in individual neurons deviate from those in the rested state. Prophylaxis of mental fatigue through short but frequent resting periods during intellectually challenging work could be theoretically justified.

\subsection*{Author contributions}
\textbf{Danko D. Georgiev:} Conceptualization, Formal analysis, Funding acquisition, Investigation, Methodology, Project administration, Resources, Software, Supervision, Validation, Visualization, Writing -- original draft, Writing -- review \& editing;\\
\textbf{Oskan B. Tasinov:} Conceptualization, Investigation, Methodology, Writing -- review \& editing;\\ 
\textbf{Danail V. Pavlov:} Conceptualization, Investigation, Methodology, Writing -- review \& editing;\\ 
\textbf{Deyan S. Hrusafov:} Conceptualization, Investigation, Methodology, Writing -- review \& editing.

\pagebreak

\subsection*{Disclosure statement}
No potential conflict of interest was reported by the authors.

\subsection*{Funding}
This study is financed by the European Union-NextGenerationEU through the National Recovery and Resilience Plan of the Republic of Bulgaria, project No. BG-RRP-2.004-0009-C02.

\subsection*{Data availability statement}
Data sharing is not applicable to this theoretical research article as no datasets were generated or analyzed during the current study.

\subsection*{Notes on contributors}
\emph{\textbf{Danko D. Georgiev}} is an Assistant Professor in Biochemistry at the Medical University of Varna, Bulgaria. Dr. Georgiev earned his M.D. from Medical University of Varna, in 2004, and his Ph.D. in Pharmaceutical Sciences from Kanazawa University, Japan, in 2008. His experience and research interests encompass topics of molecular neuroscience, molecular pharmacology, cognitive science, quantum chemistry, quantum physics, quantum information and applied mathematics.\\

\noindent\emph{\textbf{Oskan B. Tasinov}} is an Associate Professor in Biochemistry at the Medical University of Varna, Bulgaria. Dr. Tasinov earned his M.S. in Molecular Biology and Biotechnologies from Paisii Hilendarski University of Plovdiv, Bulgaria, in 2011, and his Ph.D. in Biochemistry from Medical University of Varna, in 2015. His research interests are in the molecular mechanisms of the cytoprotective, immunostimulating, anti-inflammatory, antioxidant, insulin-like and antiobesity effects of aqueous and aqueous-alcoholic extracts of medicinal plants.\\

\noindent\emph{\textbf{Danail V. Pavlov}} is a Senior Assistant Professor in Biochemistry at the Medical University of Varna, Bulgaria. Dr. Pavlov earned his M.S. in Ecology and Environmental Protection from Technical University of Varna, Bulgaria, in 2003, and his Ph.D. in Biochemistry from Medical University of Varna, in 2015. His interests include biochemistry, phytochemistry, experimental pharmacology, Black Sea ichthyofauna and quaternary palaeoecology.\\

\noindent\emph{\textbf{Deyan S. Hrusafov}} is an Associate Professor in Psychiatry and Medical Psychology at the Medical University of Varna, Bulgaria. Dr. Hrusafov earned both his degrees, M.D. in 2004 and Ph.D. in Psychiatry in 2016, from Medical University of Varna. His professional interests are in the field of general psychopathology, psychopharmacology and biological psychiatry.

\pagebreak

\end{document}